\documentclass[aps,prl,twocolumn,amsmath,amssymb,amsfonts,nofootinbib,long,floatfix]{revtex4}
\usepackage{epsfig,latexsym,bm,epstopdf}

\usepackage{color}

\makeatletter
\newcommand{\vast}{\bBigg@{4}}
\newcommand{\Vast}{\bBigg@{5}}
\makeatother
\DeclareMathAlphabet{\mathpzc}{OT1}{pzc}{m}{it}
\allowdisplaybreaks[4]
\usepackage{amsmath}
\usepackage{amssymb}

\begin{document}

\title{Detection of extremely low frequency gravitational wave using gravitational lens}

\author{Wenshuai Liu$^{1}$}
\email{674602871@qq.com}
\affiliation{$^1$School of Physics, Huazhong University of Science and Technology, Wuhan 430074, China}

\date{\today}

%***********************************   Abstract   ********************************************%

\begin{abstract}
The effect of gravitational wave of cosmological wavelength on the gravitational lensing is investigated. When the source, deflector, and observer are aligned in a highly symmetric configuration, an Einstein ring will be observed by the observer. There will be no time delays between different locations on the Einstein ring in the absence of gravitational wave. Otherwise, time delays between different locations on the Einstein ring will emerge if cosmological gravitational wave propagates through the system. Previous studies demonstrated that the time delay resulting from the aligned source-deflector-observer configuration in the presence of gravitational wave of cosmological wavelength could be equivalent to that of a similar lens with a nonaligned source-deflector-observer configuration in the absence of gravitational wave. Results in this work show that gravitational lens with the aligned source-deflector-observer configuration could serve as a potential gravitational wave detector when the whole Einstein ring observed by the observer is taken into account.
\end{abstract}

%*********************************************************************************************%

\pacs{98.80.-k,98.62.En}

%98.80.-k -> Cosmology
%98.62.En -> Electric and magnetic fields

\maketitle

%*********************************************************************************************%

\section{Introduction}

Primordial gravitational waves (PGW) with a nearly scale-invariant spectrum \cite{1,2,3,4,5,6,7} are one of the key predictions of inflation which gave rise to the structure in Universe by producing seed perturbations \cite{9,10,11,12}. Detection of PGW background is of great significance in cosmology due to the fact that information of the early Universe could be deduced based on PGW, allowing one to determine the energy scale of inflation and test different scenarios associated with the origin of the initial perturbations. The widely considered indirect method of detecting PGW with extremely low frequency in the range of $10^{-18}$Hz-$10^{-16}$Hz is the signature imprinted on cosmic microwave background (CMB) polarization pattern, the so-called B-modes of polarization\cite{13,14}. Direct detection of inflationary gravitational waves at deci-Hertz frequencies is one of the ambitious goals of spaced-based gravitational-wave detectors.

Although CMB polarization shows to be a promising probe of extremely low frequency PGW, finding a complementary observational feature induced by PGW besides its effect on the polarization of CMB will provide an alternative confirmation. Galaxy-galaxy n-point correlation functions with an upper limit on the stochastic gravitational wave background are investigated in \cite{17}. \cite{18,1801,1802} studied the proper motion of distant astrophysical objects influenced by gravitational wave. Perspective of using gravitational lensing of the CMB has been proposed by \cite{19} for the detection of such gravitational wave. Propagation of light through gravitational waves produced oscillations in apparent position of the source at the period of the gravitational wave, causing a characteristic pattern of apparent proper motions given that intervals of time is much shorter than the gravitational wave period with the condition that the frequency of the gravitational wave is limited to $\omega \gg \frac{c}{l}$ ($\omega \gg 10^{-17}$ if $l=1 Gpc$) where $l$ is the distance from the source to the observer \cite{1803}. The idea to detect gravitational waves via microlensing has been proposed by \cite{1804,1805}, where the gravitational waves are emitted by binary sources.

Another possible way to detect extremely low frequency PGW was proposed by Allen \cite{20,15}. Perturbation induced by such PGW to the metric of gravitational lens would cause additional time delays between different images besides time delays by the difference in geometric path between, and in the gravitational potential of the deflector traversed by, the different light rays of a distant quasar. Thus, Allen \cite{20,15} suggested that gravitational lenses could serve as detectors to probe such PGW using time delays between different images of a quasar. Later on, Frieman \cite{21} showed that the method proposed by Allen \cite{20,15} couldn't detect such PGW with the conclusion that the time delay induced by cosmological gravitational wave can't be observationally distinguishable from the intrinsic time delay arising from the geometry of the gravitational lens given that source and observer are equidistant from and aligned with the deflector with the assumption that $\omega L \eta \ll 1$ and $h \ll \eta$ where $L$ is the distance from the source/observer to the deflector (the speed of light is set to be $c=1$), $\eta$ is the Einstein radius, and $\omega$, h are the frequency and the dimensionless amplitude of the gravitational wave, respectively.

In order to measure the time delay due to PGW in gravitational lens, the observer should be able to separate the gravitational wave induced time delay from the one originating from the lens geometry. Results in Frieman \cite{21}, with a model of a two dimensional plane consisted of the two images of a lensed quasar, demonstrated that the time delay resulting from the lens equation for a point (or thin axially symmetric) deflector with the aligned source-deflector-observer configuration in the presence of gravitational wave of cosmological wavelength can be equivalent to that of a similar lens with a nonaligned configuration and no cosmological gravitational wave.

In this work, we study the effect of extremely low frequency PGW on the time delay in gravitational lensing with the same aligned configuration as that in Frieman \cite{21}. With such configuration, an Einstein ring will be observed by the observer. Results from this work present that the gravitational lens with the aligned source-deflector-observer configuration in the presence of extremely low frequency PGW is observationally distinguishable from a similar lens with the source out of alignment in the absence of gravitational wave, meaning that PGW could be detected by such gravitational lens. That is to say, such gravitational lens could act as a potential PGW detector.

\section{Time delay due to gravitational wave}

\begin{figure}[t!]
\begin{center}
\includegraphics[clip,width=0.5\textwidth]{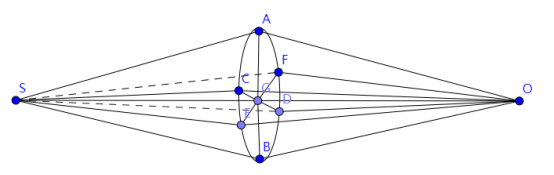}
\caption{ $S$, $G$ and $O$ represent the source, the deflector and the observer, respectively. $S$, $G$ and $O$ are on the z axis with coordinates $S$ $(x=0,y=0,z=-L)$, $G$ $(x=0,y=0,z=0)$ and $O$ $(x=0,y=0,z=L)$. $A$, $B$, $C$ and $D$ are on the Einstein ring with the condition that the line consisted of $A$, $G$ and $B$ is perpendicular to the line consisted of $C$, $G$ and $D$. $A$, $G$ and $B$ are on the x axis with coordinates $A$ $(x=L\eta,y=0,z=0)$, and $B$ $(x=-L\eta,y=0,z=0)$ while $C$, $G$ and $D$ are on the y axis with coordinates $C$ $(x=0,y=L\eta,z=0)$, and $D$ $(x=0,y=-L\eta,z=0)$.}
\end{center}
\end{figure}

We follow the same lens geometry as that in Frieman \cite{21} with the source, deflector and observer aligned in a highly symmetric configuration shown in Figure 1. A point gravitational deflector with mass $M$ is located at the origin of coordinates while the source (quasar) and the observer are at $(x=0,y=0,z=-L)$ and $(x=0,y=0,z=L)$, respectively. Given such kind of gravitational lens geometry, the observer sees an Einstein ring image of the source with an angular radius of $\eta=\sqrt{\frac{2GM}{L}}$ without time delay between different locations on the Einstein ring in the absence of gravitational wave where $\eta \ll 1$ and the speed of light is set to $c=1$. We set the gravitational potential of the deflector to be $U$ and consider a gravitational wave that propagates with an inclination angle $\phi$ with respect to the lens axis ($z$ axis). The direction of propagation of the gravitational wave lies in the $x-z$ plane. As a result, the metric induced by the gravitational wave is given as follows
\begin{eqnarray}
h_{ij}=&&
\left[\begin{array}{cccc}
    -\cos^2\phi\hspace{1 mm} h_+ &    -\cos\phi\hspace{1 mm} h_\times    & \sin\phi\hspace{1 mm}\cos\phi\hspace{1 mm} h_+ \\
    -\cos\phi\hspace{1 mm} h_\times &    h_+   & \sin\phi\hspace{1 mm} h_\times\\
    \sin\phi\hspace{1 mm}\cos\phi\hspace{1 mm} h_+ & \sin\phi\hspace{1 mm} h_\times & -\sin^2\phi\hspace{1 mm} h_+
\end{array}\right]
\nonumber \\
&& \times  \cos(\omega t-\mathbf{k} \cdot \mathbf{x})
\end{eqnarray}
where $\mathbf{k}=\omega(\sin\phi,0,\cos\phi)$ is the propagation vector, $\omega$ is the frequency, $h_+$ and $h_\times$ are the amplitude of the two polarizations of the gravitational wave, respectively.

With the above expression, we obtain the total metric given by
\begin{equation}
ds^2=(1+2U)dt^2-(1-2U)(dx^2+dy^2+dz^2)+h_{ij}dx^idx^j \label{1}
\end{equation}

Then, the time of travel of light is written as
\begin{equation}
T \approx \int_{-L}^{L}dz\left[1+\frac{1}{2}\left(\frac{dx}{dz}\right)^2+\frac{1}{2}\left(\frac{dy}{dz}\right)^2+\frac{1}{2}h_{ij}\frac{dx^i}{dz}\frac{dx^j}{dz}-2U\right] \label{2}
\end{equation}

The first three terms in the bracket of equation (\ref{2}) are due to the lens geometry while the fifth is the contribution from the gravitational potential of the deflector. The effect of gravitational wave on the time of travel is represented by the fourth term. In order to approach the level of approximation, we replace $t$ in equation (\ref{2}) by $t=t_e+(z+L)$ where $t_e$ is the time the photons were emitted at $(x=0,y=0,z=-L)$.

We study the time delay along unperturbed photon trajectories by approximating the path of photon by straight segments shown in Figure 1 with the same method in \cite{21}. Then the photon trajectories are
\begin{eqnarray}
x_{A,B}=\pm\eta(z+L), z<0 \label{3}\\
x_{A,B}=\mp\eta(z-L), z>0 \label{4}\\
y_{C,D}=\pm\eta(z+L), z<0 \label{5}\\
y_{C,D}=\mp\eta(z-L), z>0 \label{6}
\end{eqnarray}
where the subscripts represent the trajectories of photons traveling through different locations on the Einstein ring.

With the symmetry of the light paths shown in Figure 1, the only contribution to time delay is from gravitational wave represented by the fourth term in equation (\ref{2}). Integrating equation (\ref{2}) combined with equations (\ref{3})-(\ref{6}) in the case of $\phi=\frac{\pi}{2}$, $h_+\ne0$ and $h_\times=0$ for simplicity, we get the time delay with the condition of $\omega L \eta \ll 1$ and $h \ll \eta$

\begin{eqnarray}
\Delta T_{\rm AB}=&&-\frac{h\eta}{\omega}[\sin\omega(t_e+2L)+\sin\omega t_e
\nonumber \\
&&-2\sin\omega(t_e+L)\cos\omega L \eta]
\nonumber \\
&&\approx 4\frac{h\eta}{\omega}\sin^2(\frac{\omega L}{2})\sin\omega(t_e+L) \label{7}
\end{eqnarray}

\begin{eqnarray}
\Delta T_{\rm AC}=\Delta T_{\rm AD}\approx && 2\frac{h\eta}{\omega}\sin^2(\frac{\omega L}{2})\sin\omega(t_e+L)
\nonumber \\
&&-\frac{h\eta^2}{\omega}\cos\omega(t_e+L)\sin(\omega L) \label{8}
\end{eqnarray}

\begin{eqnarray}
\Delta T_{\rm CB}=\Delta T_{\rm DB}\approx && 2\frac{h\eta}{\omega}\sin^2(\frac{\omega L}{2})\sin\omega(t_e+L)
\nonumber \\
&&+\frac{h\eta^2}{\omega}\cos\omega(t_e+L)\sin(\omega L) \label{9}
\end{eqnarray}
where $h=h_+$, we define $\Delta T_{\rm MN}=T_{\rm SMO}-T_{\rm SNO}$ in which $M/N$ represents $A$, $B$, $C$ and $D$ on the Einstein ring, $T_{\rm SMO}$ represents the time travel of the light traveling from the source $S$ through the point $M$ on the Einstein ring to the observer $O$.

Based on the conclusion in Frieman \cite{21}, the time delay resulting from the lens equation for a point (or thin axially symmetric) lens with the aligned source-deflector-observer configuration in the presence of cosmological gravitational wave is equivalent to that of a lens with a nonaligned configuration with an effective misalignment angle $\beta_{\rm g}$ in the absence of gravitational wave. With such equivalent, we obtain the effective misalignment angle

\begin{equation}
\beta_{\rm g}=-\frac{h}{\omega L}\sin^2(\frac{\omega L}{2})\sin\omega(t_e+L) \label{10}
\end{equation}

The corresponding time delay between A and B is

\begin{equation}
\Delta T_{\rm AB}=-4\beta_{\rm g}\eta L = 4\frac{h\eta}{\omega}\sin^2(\frac{\omega L}{2})\sin\omega(t_e+L) \label{11}
\end{equation}

We focus on the relic gravitational wave which originated from quantum fluctuations in the early Universe and was amplified during the inflation expansion. The characteristic strain of the isotropic stochastic gravitational wave background can be written as \cite{22,23}
\begin{equation}
h_{\rm c}(f)=A\left(\frac{f}{{\rm yr}^{-1}}\right)^\alpha \label{39}
\end{equation}
where $A$, ranging from $10^{-17}$ to $10^{-15}$, is the strain amplitude at a characteristic frequency of $1$ ${\rm yr}^{-1}$. $\alpha$ ranges from $-1$ to $-0.8$.

When choosing $h=10^{-7}$, $\omega=10^{-18}{\rm Hz}$ according to equation (\ref{39}) and setting $\eta=10^{-5}$, $L=1\ {\rm Gpc}$, and assuming $\sin\omega(t_e+L)=1$, we get $\beta_{\rm g}\approx2.5\times 10^{-9}$ and
\begin{equation}
\Delta T_{\rm AB}\approx10^4 s \label{20}
\end{equation}
\begin{equation}
\Delta T_{\rm AC}=\Delta T_{\rm AD}\approx 5\times 10^3 s \label{21}
\end{equation}
\begin{equation}
\Delta T_{\rm CB}=\Delta T_{\rm DB}\approx 5\times 10^3 s \label{22}
\end{equation}

The perpendicular distance between the equivalent misaligned source and the line consisted of $G$ and $O$ is $L_{\rm mis}=2L|\beta_{\rm g}|\approx1.5\times 10^{17}m\approx5.8\times 10^{3}$ light days which is much larger than the typical physical size of a quasar (light hours to light days) \cite{24}, thus, a complete Einstein ring wouldn't be observed by the observer if there indeed exists such misalignment in the absence of gravitational wave, meaning that the gravitational lens with the aligned source-deflector-observer configuration in the presence of gravitational wave of cosmological wavelength couldn't be equivalent to a similar lens with a nonaligned configuration and no gravitational wave.

Especially, when $\sin\omega(t_e+L)=0$ and $\cos\omega(t_e+L)=1$, we get

\begin{equation}
\Delta T_{\rm AB}=0 \label{12}
\end{equation}

\begin{equation}
\Delta T_{\rm AC}=\Delta T_{\rm AD}\approx -\frac{h\eta^2}{\omega}\sin(\omega L)= -\Delta T_{\rm CB}=-\Delta T_{\rm DB} \label{13}
\end{equation}

If setting $\eta=10^{-5}$, $h=10^{-7}$, $\omega=10^{-18}{\rm Hz}$ and $L=1\ {\rm Gpc}$, we obtain $\Delta T_{\rm CA}=\Delta T_{\rm DA}=\Delta T_{\rm CB}=\Delta T_{\rm DB}\approx 1s$.

When $\phi=0$, we get the follows based on equations (\ref{1})-(\ref{6})
\begin{equation}
\Delta T_{\rm AB}=0
\end{equation}

\begin{equation}
\Delta T_{\rm AC}=\Delta T_{\rm AD}=\Delta T_{\rm BC}=\Delta T_{\rm BD}=-2Lh\eta^2 \cos\omega(t_e+L)
\end{equation}

It shows that $\Delta T_{\rm AC}=2s$ with parameters of $\eta=10^{-5}$, $h=10^{-7}$, $\omega=10^{-18}{\rm Hz}$ and $L=1\ {\rm Gpc}$ when $\cos\omega(t_e+L)=-1$.

The time delay between arbitrary two images with locations which are symmetrical about the gravitational lens is

\begin{equation}
\Delta T \approx -4\beta_g \eta L = 4\frac{h_+\eta}{\omega} \cos\theta\sin^2\left(\frac{\omega L}{2}\right)\sin\omega(t_e+L) \label{23}
\end{equation}
where $\theta$ is the angle between line consisted of the two images and the projection of the direction of propagation of gravitational wave on the plane containing the Einstein ring, and

\begin{equation}
\beta_{\rm g}=-\frac{h_+}{\omega L}\cos\theta\sin^2\left(\frac{\omega L}{2}\right)\sin\omega(t_e+L)\label{24}
\end{equation}

Throughout the above, we adopt the condition of $\omega L \eta \ll 1$ and $h \ll \eta$, and assume the same constant deflection of each light ray as the way adopted by \cite{21}. In reality, photon trajectories would be perturbed by gravitational wave \cite{21}, but the resulting time delays are consistent with equation (\ref{23}) in the limit of $\omega L \eta \ll 1$ and $h \ll \eta$ \cite{21}. The image angular positions are as follows when perturbation from gravitational wave is taken into account \cite{21}

\begin{equation}
\theta_{1,2}\approx \pm\eta+\frac{1}{2}\beta_g \label{19}
\end{equation}

With the parameters above equation (\ref{20}) and combined with equation (\ref{24}) and (\ref{19}), it shows that $\beta_{\rm g} \ll \eta$, meaning that the relative deflection $\frac{1}{2}\beta_g$ is neglectable compared with $\eta$. Thus, it is expected that a full Einstein ring could be observed by the observer, resulting that the difference of fluxes/magnifications between specific images on the Einstein ring in the presence of gravitational wave is neglectable since propagation of light through gravitational waves preserves surface brightness and total intensity of a source \cite{2301,2302,2303}. Therefore, the conclusion that results derived above with assumption of the same constant deflection of each light ray is equivalent to those obtained from the real photon trajectories perturbed by gravitational wave, is valid only when the condition of $\omega L \eta \ll 1$ and $h \ll \eta$ is satisfied (in fact, the condition of $\omega L \eta \ll 1$ and $h \ll \eta$ could be satisfied easily in the case of gravitational wave of cosmological wavelength).

In the general case of an aligned source-deflector-observer configuration with a gravitational wave of arbitrary polarization and direction of propagation, the time delay between arbitrary two images with locations which are symmetrical about the gravitational lens is

\begin{eqnarray}
\Delta T &\approx& -4\beta_g \eta L = \frac{4\eta}{\omega}\sin\phi\left[h_+\cos\theta+h_\times\frac{2\sin\theta}{1-\cos\phi}\right]
\nonumber \\
&\times& \sin^2\left[\frac{\omega L}{2}(1-\cos\phi)\right]\sin\omega(t_e+L) \label{16}
\end{eqnarray}
where
\begin{eqnarray}
\beta_{\rm g}&=&-\frac{1}{\omega L}\sin\phi\left[h_+\cos\theta+h_\times\frac{2\sin\theta}{1-\cos\phi}\right]
\nonumber \\
&\times& \sin^2\left[\frac{\omega L}{2}(1-\cos\phi)\right]\sin\omega(t_e+L)
\end{eqnarray}

Let's reconsider the lens geometry shown in Figure 1. Assuming that the angle between the line AGB and the projection of the direction of propagation of gravitational wave on the plane containing the Einstein ring is $\theta$, the angle between the line AGB and the line CGD is $\angle AGC = a$ ($a$ is an arbitrary angle and not limited to $\frac{\pi}{2}$ in this case) and the angle between the line CGD and the line EGF is $\angle CGE = b$, we obtain the following relationship between $\Delta T_{AB}$, $\Delta T_{CD}$ and $\Delta T_{EF}$ based on equation (\ref{16})

\begin{equation}
\Delta T_{CD}=\Delta T_{AB}\left[\cos a-\frac{\sin a \cos(a+b)}{\sin(a+b)}\right] +\Delta T_{EF}\frac{\sin a }{\sin(a+b)} \label{17}
\end{equation}
which is an unique feature possessed only by an aligned source-deflector-observer configuration in the presence of gravitational wave with extremely low frequency and where $\sin(a+b)\ne 0$.

From the results in \cite{25}, it is expected that an aligned lens in the presence of large scale structure fluctuations is equivalent to a nonaligned lens with an effective misalignment angle in the absence of perturbations from large scale structure. But the large scale structure can't produce the particular relationship shown in equation (\ref{17}), meaning that large scale structure perturbations couldn't mimic the time delays resulting from cosmological gravitational waves and couldn't rule out the possibility of using lensed quasars as gravitational wave detectors.

\section{Conclusion and discussion}
We study the time delays in an aligned lens configuration where the source and the observer are equidistant from the deflector. In contrast to the conclusion from Frieman \cite{21} where the time delay resulting from the lens equation for a point (or thin axially symmetric) deflector with the aligned source-deflector-observer configuration in the presence of gravitational wave of cosmological wavelength could be equivalent to that of a lens with the source out of alignment in the absence of gravitational wave, results of this work show that gravitational lens where the source, lens and observer are aligned could serve as a possible detector of cosmological gravitational wave.

In order to use gravitational lens as a gravitational wave detector, the observer should be able to separate the time delay induced by gravitational wave from the intrinsic time delay due to the lens geometry. When only considering the time delay between the images in the two dimensional plane consisted of the two images and the aligned source-deflector-observer, the observer couldn't distinguish the origin of the time delay due to the fact that the lens geometry is not a priori and should be deduced from detailed observations, resulting that the observer can't tell whether the time delay originates from the gravitational wave in the aligned source-deflector-observer geometry or it is a natural outcome of the gravitational lens with a small misalignment in the absence of gravitational wave. When the whole Einstein ring is taken into account, images from different locations on the Einstein ring have different time travel with relationship shown in equation (\ref{17}), meaning that such gravitational lens could be used as a possible detector to confirm the presence of gravitational wave with extremely low frequency.

Up to now, only dozens of Einstein ring systems have been discovered \cite{26,27,28}, showing complete or nearly complete ring morphology. In future, many gravitational lenses with nearly perfect Einstein rings of distant quasars may be discovered \cite{29,30}. Quasars usually exhibit variability originating from the inner accretion disk surrounding the central supermassive black hole and the time resolution in observation could approach about minutes \cite{31,32} which is far less than the time delay due to cosmological gravitational wave from inflation shown in equation (\ref{16}). Thus, it is possible to use gravitational lens with an aligned source-deflector-observer configuration to detect extremely low frequency gravitational wave.


\begin{thebibliography}{99}



\bibitem{1}L. F. Abbott \& M. B. Wise, Nucl. Phys. B244, 541 (1984).
\bibitem{2}A. Starobinskii, Sov. Astron. Lett. 11, 133 (1985).
\bibitem{3}V. A. Rubakov, M.V. Sazhin, \& A.V. Veryaskin, Phys. Lett. 115B, 189 (1982).

\bibitem{4}R. Fabbri \& M. D. Pollock, Phys. Lett. 125B, 445 (1983).
\bibitem{5}A. A. Starobinsky, JETP Lett. 30, 682 (1979).
\bibitem{6}V. Sahni, Phys. Rev. D 42, 453 (1990).
\bibitem{7}B. Allen, Phys. Rev. D 37, 2078 (1988).


\bibitem{9}L. P. Grishchuk, JETP Lett. 23, 293 (1976).

\bibitem{10}L. P. Grishchuk, Sov. Phys. Usp. 20, 319 (1977).

\bibitem{11}A. A. Starobinsky, Phys. Lett. B 91, 99 (1980).


\bibitem{12}A. D. Linde, Phys. Lett. B 108, 389 (1982).


%\bibitem{13}L. F. Abbott, \& M. B. Wise, Nucl. Phys. B, 244, 541 (1984)
%\bibitem{14}A. A. Starobinsky, Sov. Astron. Lett., 11, 133 (1985)
%\bibitem{15}V. A. Rubakov, M. V. Sazhin, \& A. V. Veryaskin, Phys. Lett., 115B, 189 (1982)
%\bibitem{16}R. Fabbri \& M. D. Pollock, Phys. Lett. 125B, 445 (1983)

\bibitem{13}M. Kamionkowski, A. Kosowsky, \& A. Stebbins, Phys. Rev. Lett. 78, 2058 (1997).

\bibitem{14}U. Seljak, \& M. Zaldarriaga, Phys. Rev. Lett. 78, 2054 (1997).





\bibitem{17}E. V. Linder, Astrophys. J., 328, 77 (1988).

\bibitem{18}R. Bar-Kana, Phys. Rev. D, 54, 7138 (1996).

\bibitem{1801}L. G. Book, \& {\'E}. {\'E}. Flanagan, Phys. Rev. D 83, 024024 (2011).

\bibitem{1802}S. Bharadwaj, \& T. Guha Sarkar, Phys. Rev. D, 79, 124003 (2009).

\bibitem{1803}T. Pyne, C. R. Gwinn, M. Birkinshaw, T. M. Eubanks, \&  D. N. Matsakis, Astrophys. J., 465, 566 (1996).




\bibitem{1804}S. L. Larson, \& R. Schild, ArXiv e-prints, arXiv:0007142 (2000).

\bibitem{1805}R. Ragazzoni, G. Valente, \& E. Marchetti, Mon. Not. R. Astron. Soc. 345, 100 (2003).



\bibitem{19}L. G. Book, M. Kamionkowski, \& T. Souradeep, Phys. Rev. D, 85, 023010 (2012).

\bibitem{20}B. Allen, Phys. Rev. Lett., 63, 2017 (1989).

\bibitem{15}B. Allen, Gen. Relativ. Gravit., 22, 1447 (1990).



\bibitem{21}J. A. Frieman, D. D. Harari, \& G. C. Surpi, Phys. Rev. D, 50, 4895 (1994).



\bibitem{22}L. P. Grishchuk, Phys. Uspekhi, 48, 1235 (2005).

\bibitem{23}F. A. Jenet, et al., Astrophys. J., 653, 1571 (2006).

\bibitem{2301}H. Bondi, F. A. E. Pirani, \& I. Robinson, Proc. R. Soc. London A, 251, 519 (1959).
\bibitem{2302}D.M. Zipoy, Phys. Rev., 142, 825 (1966)

\bibitem{2303}R. Penrose, in Perspectives in Geometry and Relativity, ed. B. Hoffmann (Bloomington: Indiana Univ. Press), 259 (1966)




\bibitem{24}R. C. Hickox, \& D. M. Alexander, ARAA, 56, 625 (2018).

\bibitem{25}G. C. Surpi, D. D. Harari, \& J. A. Frieman, Astrophys. J., 464, 54 (1996).


\bibitem{26}A. S. Bolton, et al., Astrophys. J., 682, 964 (2008).

\bibitem{27}A. S. Bolton, et al., Astrophys. J., 684, 248 (2008).
\bibitem{28}D. P. Stark, et al., Mon. Not. R. Astron. Soc., 436, 1040 (2013).


\bibitem{29}C.-H. Lee, PASA, 34, e014 (2017).
\bibitem{30}C. Avestruz, et al., Astrophys. J, 877, 58 (2019).

\bibitem{31}S. AL Otaibi, PhD thesis, Machine learning methods for delay estimation in gravitationally lensed signals (2016).

\bibitem{32}A. G. de Bruyn, \& J.-P. Macquart, Astron. Astrophys., 574, A125 (2015).











\end{thebibliography}
\end{document}